\documentstyle[12pt,epsf]{article}

\def\beq{\begin{equation}}
\def\eeq{\end{equation}}
\def\beqa{\begin{eqnarray}}
\def\eeqa{\end{eqnarray}}
\def\a {{\rm f}}

\def\g{\xi}
\def\w{\rho}
\def\y{\eta}

\begin{document}

\begin{flushright}
ITP-SB-98-46
\end{flushright}

\begin{center}{\bf\Large\sc Energy and Color Flow in Dijet Rapidity Gaps}
\vglue 1.2cm
\begin{sc}
Gianluca Oderda and George Sterman\\
\vglue 0.5cm
\end{sc}
{\it Institute for Theoretical Physics \\
SUNY at Stony Brook,
Stony Brook, NY 11794-3840, USA}\\
\end{center}
\vglue 1cm
\begin{abstract}
When rapidity gaps in high-$p_T$ dijet
events are identified by energy flow in the central
region, they may be calculated from factorized
cross sections in perturbative QCD, up to corrections that
behave as inverse powers of the central region energy.
Although power-suppressed corrections may be important,
a perturbative calculation of dijet rapidity gaps in
${\rm p}\bar{\rm p}$ scattering successfully 
reproduces the overall features observed at the Tevatron.
In this formulation, the average color content of
the hard scattering is well-defined.
We find that hard dijet rapidity gaps in quark-antiquark scattering
are not due to singlet exchange alone.  
\end{abstract}

\centerline{PACS Nos.: 12.38.Aw, 12.38.Cy, 13.85.-t, 13.87.-a} 

\newpage

Among the most intriguing recent experimental results in
quantum chromodynamics is the observation 
of dijet rapidity-gap events, with anomalously low radiation
in a wide interjet rapidity region \cite{D0,CDF,ZEUS}.
These events are typically identified 
by low or zero hadron multiplicity in the central region, 
despite the high momentum transfer
necessary to produce the jets.  

The existence of such events was originally suggested on the basis of
color flow considerations in QCD \cite{DKT,Bjork}.  If forward jets are produced 
by exchanging a pair of gluons in a color singlet state, color can be
recombined independently in each forward region.  Then
much less radiation is expected between the jets than when the exchange is a color octet
gluon, which requires recombining color between particles moving
in nearly opposite directions.  Rapidity gap events have special interest
as clear illustrations of color coherence 
and its interplay with hadronization.
In addition, because their observation requires large rapidity intervals, they offer 
a new window into a perturbative, yet Regge-like limit of QCD. 
Nonetheless, despite their intuitive appeal, the theoretical understanding of rapidity gaps has
been somewhat hampered by two problems. One of these is
the issue of ``survival" \cite{Bjork,GLM}.   In any
high-energy scattering, multiple soft interactions between spectators
of the hard interaction may fill the gap by processes unrelated
to the color content of the hard interaction. 
The second is 
that, since even the softest gluon carries color in the octet representation,
it is not immediately obvious how the color of the hard scattering
is to be defined.

In this paper, we observe that it is possible to overcome these
problems, at least in part, by identifying rapidity gaps in terms of energy flow, rather
than multiplicity.  The energy flow $Q_c$ into the central rapidity interval between
a pair of jets is an infrared safe observable.  That is, $d\sigma/dQ_c$ can
be written as a convolution of parton distributions with a
perturbative hard-scattering function, which depends on $Q_c$.  
The issue of color flow may then be formulated self-consistently
in the hard-scattering function.   Corrections
to the factorized cross section are proportional to powers of $\Lambda/Q_c$,
with $\Lambda$ the scale of
the QCD coupling, and may become large for small $Q_c$.  As we shall see, however,
the purely perturbative cross section remains well-defined,
and  energy flow gaps appear in this
limit, once soft radiation is resummed
including color effects \cite{BottsSt,KOS1,KOS2}.  
In the conventional formulation for rapidity gaps, one writes
 $f_{\rm gap}=f_{\rm singlet}P_S$, with $f_{\rm gap}$ the fraction of
gap events, $f_{\rm singlet}$ the fraction of ``hard singlet" exchanges, 
and $P_S$ the
survival probability.  Compared to this, we generalize $f_{\rm singlet}$, which then
necessarily incorporates a perturbative survival probability.
Nonperturbative survival considerations may 
reappear as we approach zero energy in the gap, but their importance should be
reduced in a calorimetric measurement.  Our results below support this possibility. 

To be specific, we will study the process
$p(p_A)+\bar{p}(p_B) \rightarrow J_1(p_1)+J_2(p_2)+X_{{\rm gap}}$,
where we sum inclusively over final states, while measuring the
energy that flows into
the intermediate region between two forward jets.
For simplicity, we restrict ourselves to valence quarks and
antiquarks, $q(k_A)+\bar{q}(k_B) \rightarrow q(k_1)+\bar{q}(k_2)+X$.
We will begin by deriving a cross section 
for this process, specific to the geometry described by
the D0 and CDF collaborations \cite{D0,CDF,D0fig}.
We go on to evaluate the cross section as a function of $Q_c$,
and to view the results in the light of what we have learned from experiment.
We will close with a few comments on the relation
of our approach to previous work, and on prospects for further
development in this problem.

\begin{figure}
\centerline{\epsffile{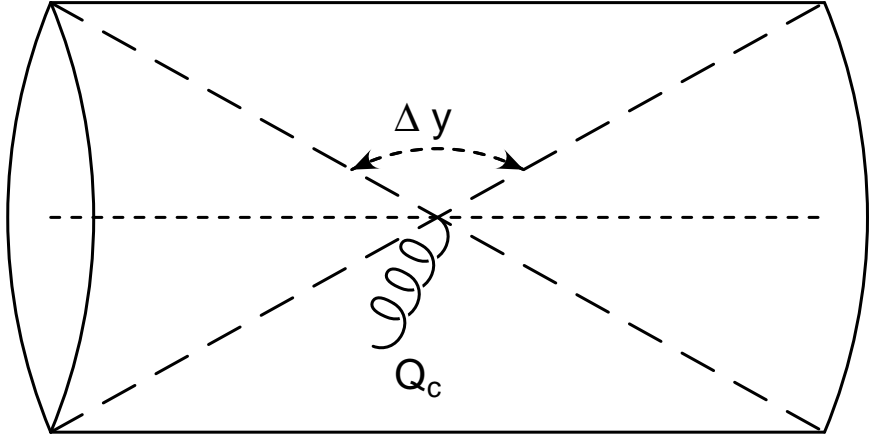}}
\caption[dum]{\small{
Geometry of the calorimeter detector. $Q_c$ is the energy flow into the central
rapidity interval
}}
\label{geometry}
\end{figure}

Following CDF and D0, we require 
the two jets, and therefore the outgoing
partons coming from the hard scattering, $q(k_1)$, $\bar{q}(k_2)$, 
to be directed into fixed forward and backward (collectively denoted ``forward") regions of 
the calorimeter, defined by 
$|y|>y_0$, where $y$ is the (pseudo)rapidity $y=(1/2)\ln\cot(\theta/2)$,
with $\theta$ the polar angle.  In addition we require the jets
to have transverse energies above an experimental threshold, $E_T$. 
We will discuss cross sections for measured 
energy in a symmetric central region, spanning rapidity 
$\Delta y=2y_0$. This geometry is presented schematically in Fig.\ 
\ref{geometry}.
The inclusive dijet cross section for all events with energy in
the central region equal to $Q_c$ is a typical factorizable
jet cross section, which may be written as
\beqa
&&\frac{d\sigma}{dQ_c}\left(S,E_{T},\Delta y\right) 
=
 \sum_{f_A,f_B=u,d} \int d\cos\hat{\theta}\,
\nonumber\\ 
&& \times \hspace{5mm} \int_0^1 dx_A \int_0^1 dx_B\;
 \phi_{f_A/p}(x_A,-\hat{t}) \,
\phi_{\bar{f}_B/\bar{p}}(x_B,-\hat{t}) \nonumber \\
&& \hspace {10mm} \times \sum_{f_1,f_2=u,d}\frac{d\hat{\sigma}^{(\a)}}
{dQ_c \, d\cos\hat{\theta}} 
\left(\hat{t},\hat{s},y_{JJ},\Delta y,\alpha_s(\hat{t})\right)\, ,
\nonumber \\
\label{crosssec}
\eeqa
where
$\phi_{f_A/p}$, $\phi_{\bar{f}_B/\bar{p}}$ are 
valence parton distributions, evaluated at scale $-\hat t$,
the dijet momentum transfer. ${d\hat{\sigma}^{(\a)}}/{dQ_c \, 
d\cos\hat{\theta}}$
is a hard scattering function, starting with the Born cross section
at lowest order.  The 
index $\a$ denotes
$f_A+\bar{f}_B \rightarrow f_1 +\bar{f}_2$. 
The detector geometry determines the phase space
for the dijet total rapidity, $y_{JJ}$, 
the partonic center-of-mass (c.m.)\ energy squared, $\hat{s}$, and
the partonic c.m.\ scattering angle $\hat{\theta}$, with
$-\frac{\hat{s}}{2}\left(1-\cos \hat{\theta} \right)=\hat{t}$.
For simplicity of presentation, we take $Q_c$ to be the energy in
the dijet c.m.\ .

In the spirit of Refs.\
\cite{BottsSt,KOS1,KOS2}, we now observe that we may perform
a further factorization on 
the partonic hard-scattering  function  ${d\hat{\sigma}^{(\a)}}/{dQ_c \, 
d\cos\hat{\theta}}$.
The underlying observation is that for $Q_c\ll \sqrt{-\hat{t}}$,
the soft gluon radiation that appears in the central region 
decouples from the dynamics of the hard interaction that produces the dijet
event.  In technical terms, soft gluon emission may
be approximated by an effective cross section, in which the
hard scattering is replaced by a product of recoilless color sources
(specifically, Wilson lines \cite{KOS1,KOS2}) in the directions
of the incoming partons and the outgoing jets.  
The refactorized hard-scattering function then takes the form \cite{StOd}
\beqa
Q_c\frac{d\hat{\sigma}^{(\a)}}{dQ_c}\left(\hat{s},\hat{t},y_{JJ},\Delta y,\alpha_s(-\hat t)\right)
&=&
H_{IL}\left( \frac{\sqrt{-\hat{t}}}{\mu},
\sqrt{\hat{s}},\sqrt{-\hat t},\alpha_s(\mu^2) \right) \nonumber \\
&& \hspace{-10mm} \times S_{LI} \left( \frac{Q_c}{\mu},y_{JJ},\Delta y\right).
\label{factor}
\eeqa
The functions $S_{LI}$ and $H_{IL}$ contain the dynamics of
soft radiation from Wilson lines (at measured $Q_c)$,
and the hard interaction, respectively, with $\mu$ a new
factorization scale.  The product itself must be independent
of how we choose the scale, so long as $\mu>Q_c$.
The indices $I$ and $L$ label the possible color structures of
the hard interaction, which correspond in $S_{LI}$ to the
color matrices that couple the four Wilson lines
representing the $2\rightarrow 2$ hard subprocess.  One index
refers to the hard scattering in the amplitude, the other to the
hard scattering in the complex conjugate.  
For quark-antiquark scattering, the Wilson lines are in the 3 (quark) and 3$^*$ 
(antiquark) representations
of $SU(3)$, respectively, and their product may be characterized
by either singlet or octet color exchange in the $t$- or $s$-channel.
For the physical reasons outlined above, we will choose a $t$-channel color basis.
Corrections to Eq.\ (\ref{factor}) are expected from
{\it three-jet} final states, for which the analysis below must, in principle,
be repeated.  We expect these corrections to be relatively small.

Because the left-hand side of  Eq.\ (\ref{factor}) is independent of 
the precise choice of factorization scale $\mu$,  the matrices $H_{IL}$
and $S_{LI}$ must satisfy evolution equations, in which their
variations with $\mu$ cancel each other.  The only 
variable that $H$ and $S$ hold  in common is $\alpha_s(\mu^2)$, and,
as a result, the evolution
equation for $S$ is 
\beq 
\left(\mu\frac{\partial}{\partial\mu}+\beta(g)\frac{\partial}{{\partial}g}
\right)S_{LI}=
-(\Gamma_S^\dagger)_{LB}S_{BI}-S_{LA}(\Gamma_S)_{AI}\, ,
\label{eq:resoft}
\eeq
and similarly for $H$, with $\Gamma_S(\alpha_s)$ an anomalous dimension
matrix.
Consider a $t$-channel singlet-octet basis,
with color vertices schematically given by
\beq
c_1=I\times I\, ,\ c_2=\sum_a T^a\times T^a\, ,
\label{eq:basqqbar}
\eeq
with $I$ the identity and $T^a$ the generators of $SU(3)$ in the quark representation.
A one-loop calculation in this basis gives \cite{StOd}
\beq
\Gamma_S\left(y_{JJ},\Delta y,\hat{\theta}\right)
=
\frac{\alpha_s}{4\pi} \left(
\begin{array}{cc}
\w+\g & -4\frac{C_F}{N_c}i\pi \\
-8i\pi & \w-\g
\end{array} \right),
\label{rapanodim}
\eeq
where $N_c$ is the number of colors.
The functions $\g$ and $\w$ are
\beqa
\g(\Delta y)&=&-2N_c \Delta y +2i\pi\frac{N_c^2-2}{N_c},
\label{defntg}
\\
&&\hspace{-10mm}\w(y_{JJ},
\Delta y,\hat{\theta}
)=\frac{N_c^2-1}{N_c} \nonumber \\
&& \times \left[ \ln \left( \frac{\cos(\hat{\theta})+
\tanh\left( \frac{\Delta y}{2}-y_{JJ}\right)}
{\cos(\hat{\theta})-\tanh\left( \frac{\Delta y}{2}-y_{JJ} \right)}\right) \right.
\nonumber \\
&&+ \left. \ln\left(\frac{\cos(\hat{\theta})-\tanh\left( -\frac{\Delta y}{2}-y_{JJ} 
\right)}
{\cos(\hat{\theta})
+\tanh\left( -\frac{\Delta y}{2}-y_{JJ} \right)}\right)\right] \nonumber \\ 
& & +\frac{2}{N_c}\Delta y-2i\pi\frac{N_c^2-2}{N_c}\, .
\label{defntw}
\eeqa
While $\rho$ depends
on the jet rapidities 
and on the partonic scattering angle,  
$\xi$ depends on the geometry only,
through $\Delta y$.   We note that the off-diagonal
components of $\Gamma_S$ are purely imaginary.  This 
interesting feature is due to strong coherence effects in the one-loop calculation,
related to angular ordering \cite{cohere}.

To study the $Q_c$-dependence of $S$,
it is convenient to diagonalize $\Gamma_S$. 
In the basis in which $\Gamma_S$ is diagonal, Eq.\ (\ref{eq:resoft})
implies that the components of $S$ evolve independently in $\mu$.   
In this basis we may calculate unambiguously
the dependence on the central energy flow $Q_c$.
This is the technique that we summarize in the following.

The eigenvectors of $\Gamma_S$ in Eq.\ (\ref{rapanodim}) may be chosen as
\beqa
e_1&=&\left(\begin{array}{c}
1\\
\frac{8\pi}{i}\left(\g-\frac{1}{\sqrt{N_c}}\y \right)^{-1} \\
\end{array} \right) \nonumber \\
e_2&=&\left(\begin{array}{c}
\frac{i}{8\pi}\left(\g+\frac{1}{\sqrt{N_c}}\y \right)\, , \\
1\\
\end{array} \right),
\label{eigenvectors}
\eeqa
where we define
\beq
\y(\Delta y)\equiv\sqrt{N_c \left[\g(\Delta y)\right]^2 -32C_F\pi^2}\, .
\label{ydef}
\eeq
A very useful feature is that these eigenvectors are
independent of the jet rapidities, and depend only on
$\Delta y$.
The corresponding eigenvalues of $\Gamma_S$ are in general complex,
\beqa
\lambda_1&=&\frac{\alpha_s}{2\pi}\left[\frac{1}{2}\, \w-
\frac{1}{2\sqrt{N_c}} \y \right] 
\nonumber \\
\lambda_2&=&\frac{\alpha_s}{2\pi}\left[\frac{1}{2}\w+
\frac{1}{2\sqrt{N_c}} \y \right] .
\label{eigenvalues}
\eeqa
In the limit of a large central region, $\Delta y \gg 1$, the function $\g$ 
has a large negative real part, while the real part of $\eta$ is
positive, and grows with $\Delta y$.
 From Eq.\ (\ref{eigenvectors}), we see that, as $\Delta y\rightarrow \infty$, $e_1$ 
reduces to a color 
 ``quasi-singlet'', and 
$e_2$ to a color ``quasi-octet''. As a realistic
 example, we take the value 
$\Delta y=4$, and find
\beqa
e_1&=&\left(\begin{array}{c}
1\\
0.455 \, e^{2.161 \, i} \\
\end{array} \right) \nonumber \\
e_2&=&\left(\begin{array}{c}
0.101 \, e^{-0.981 \, i} \\
1\\
\end{array} \right).
\label{eigenvnumb}
\eeqa 
For this configuration, the second eigenvector is close to a color octet,
but the first is still a mixture of octet and singlet, with the
latter only slightly predominant. In the following, however, we find it suggestive to retain 
the names ``quasi-singlet'' and ``quasi-octet'' for the elements of the diagonal basis.  
In the limit of large $\Delta y$,
the eigenvalue for the quasi-octet grows with $\Delta y$, while
the eigenvalue of the quasi-singlet does not.  This will produce
the expected enhancement of the latter relative to the former in the resummed cross section at small $Q_c$.
We shall use Greek indices to identify the basis in which  $\Gamma_S$ is diagonal.

We can now write down a resummed cross section, working to lowest
order in $\alpha_s(-\hat{t})$, but resumming all leading logarithms in $Q_c$.
We transform Eq.\ (\ref{factor}) to the diagonal basis, and solve
the evolution equation for $S$, to get
\beqa
\frac{d\hat{\sigma}^{(\a)}}{dQ_c \, d\cos \hat{\theta}}\left(\hat{s},\hat{t},y_{JJ},\Delta y,\alpha_s(-\hat t)\right)
&=& 
\nonumber \\
&& \hspace{-40mm} H^{(1)}_{\beta \gamma}\left( \Delta y,\sqrt{\hat{s}},\sqrt{-\hat{t}},
\alpha_s\left(-\hat{t}\right) \right) 
 S^{{(0)}}_{\gamma \beta} ( \Delta y ) \,  \nonumber \\
&& \hspace{-40mm} \times {E_{\gamma\beta} \over Q_c}\; 
\left[\ln\left({Q_c\over \Lambda}\right)\right]^{E_{\gamma\beta}-1}\;  
\left[ \ln \left( {\sqrt{-\hat{t}}\over \Lambda}\right)\right]^{-E_{\gamma\beta}}\, . \nonumber\\
\label{factor2}
\eeqa
The coefficients  $E_{\gamma \beta}$ are given by
\beqa
&&E_{\gamma\beta}\left(y_{JJ},\hat{\theta},\Delta y \right)
=\frac{2\pi}{\beta_1}\, \left[\hat{ \lambda}^{*}_{\gamma} \left(y_{JJ},
\hat{\theta},
\Delta y \right) +\hat{ \lambda}_{\beta} \left( y_{JJ},\hat{\theta},
\Delta y \right) \right]\, ,
\label{expon}
\eeqa
where $\beta_1$ is the first coefficient in the expansion of the QCD
$\beta$-function, $\beta_1=\frac{11}{3}N_c-\frac{2}{3}n_f$, 
and where we define $\hat{ \lambda}_{\beta}$ by 
$\lambda_{\beta}=\alpha_s \hat \lambda_\beta+\cdots$.

In accordance with our approximation,
the matrix $S^{{(0)}}_{\gamma \beta}$ 
is obtained by transforming the zeroth order $S^{{(0)}}_{LI}$ of 
Eq.\ (\ref{factor})
to the new basis.
The matrix $S^{{(0)}}_{LI}$ is just a set of color traces,
\beq
S^{{(0)}}_{LI}=\left(\begin{array}{cc}
N^2_c & 0\\
0 & \frac{1}{4} \left( N^2_c-1 \right) 
\end{array} \right),
\label{softcol}
\eeq
and is transformed to the diagonal basis by the 
matrix 
${\left( R^{-1} \right)}_{K \beta} \equiv \left( {e_{\beta}}
\right)_K$ \cite{KOS2},
\beq
S^{(0)}_{\gamma \beta} \equiv 
\left[ \left( R^{-1}\right)^\dagger \right]_{\gamma M}
S^{(0)}_{MN}
{\left( R^{-1} \right)}_{N \beta}\, .
\label{eq:newbasS}
\eeq
Analogously, we take for $H^{(1)}_{IL}$ the square of
the single-gluon exchange amplitude, represented 
in the color basis.
Considering the dominant $t$-channel Born-level amplitude  alone,
which is purely octet, we have
$H^{(1)}_{IL}=\delta_{I2}\, \delta_{L2} \, \hat{\sigma}_t$,
where $\hat{\sigma}_t$ is the  $t$-channel 
partonic cross section, including 
the coupling $\alpha_s(-\hat{t})$.  
The contribution  of $s$-channel  diagrams
has a relatively small effect, and will be described elsewhere \cite{StOd}.
In the diagonal basis the hard matrix $H^{(1)}_{IL}$ becomes   
$H^{(1)}_{\beta\gamma}$, defined as
\beq
H^{(1)}_{\beta\gamma}={\left( R \right)}_{ \beta L} \, 
H^{(1)}_{LK}
\, {\left( R^{\dagger} \right)}_{K \gamma}.
\label{newbasH}
\eeq 
Observe that  $S^{(0)}$ and  $H^{(1)}$ 
both acquire a $\Delta y$-dependence through the change of basis.

\begin{figure}
\centerline{\epsffile{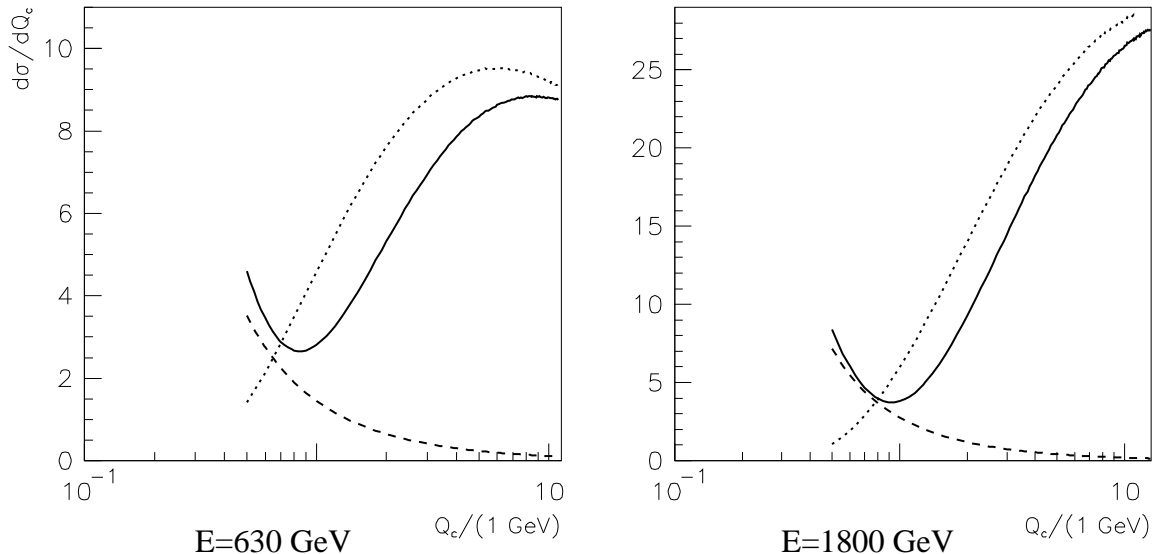}}
\caption[dum]{\small{
The cross section (solid line) and the contributions from
quasi-octet (dotted line) and quasi-singlet (dashed line), for
$\sqrt{S}=630$ GeV, $\Delta y=3.2$,
and  $\sqrt{S}=1800$ GeV, $\Delta y=4.0$, respectively.
Compare Fig.\ 1 of Ref.\ \cite{D0fig}.  Units are arbitrary.
}}
\label{results}
\end{figure}

From Eqs.\ (\ref{factor2})-(\ref{expon}), using the results described above, it is possible 
to evaluate Eq.\ (\ref{crosssec}).
For the valence partons we have taken the leading order 
CTEQ4L distributions \cite{cteq4}.
In Fig.\ \ref{results} we plot the shapes of the cross sections obtained 
in this way,
as a function of the radiation into the  central region, $Q_c$,
for two different
sets of conditions, $\sqrt{S}=630$ GeV, $\Delta y=3.2$,
 and  $\sqrt{S}=1800$ GeV, $\Delta y=4.0$.
We also show the contributions of quasi-octet and quasi-singlet terms.
There is in addition a negative interference term, not exhibited separately.
As anticipated above, we find a strong suppression   
of the quasi-octet component for very small values of $Q_c$, contrasted to a peak for
the quasi-singlet in this limit. The reason for this difference is easily
found. For most kinematic configurations, 
the coefficient $E_{11}$ from Eq.\ (\ref{expon}) is less than one,
so that the quasi-singlet cross section in Eq.\ (\ref{factor2}) decreases monotonically
with increasing  $Q_c$.  For the quasi-octet,
on the other hand, $E_{22}$ is always greater than unity, so that its contribution
grows with $Q_c$, until the power of the logarithm is overcome by
the dimensional factor of $1/Q_c$.  

These results can be compared with the experimental data
in Fig.\ 1 of Ref.\ \cite{D0fig}, showing the measured number of events as a function of
the number of
towers counted in the central region of the calorimeter, clearly related to $Q_c$.
We can understand the similarity of shapes in terms of the 
$Q_c$ dependence in Eq.\ (\ref{factor2}), discussed above.
This similarity is suggestive; indeed,  
from our simulation we have 
evaluated the minimum-maximum ratio of the cross section, finding about $30\%$ at 
$\sqrt{S}=630 \, {\rm GeV}$ and about $15 \%$ at $\sqrt{S}=1800 \, {\rm GeV}$, 
close to the analogous ratios in  Fig.\ 1 of Ref.\ \cite{D0fig}.
We have also determined an analog of a ``hard singlet fraction"
\cite{D0,CDF}, as the ratio of the area under the quasi-singlet curve
to the area under the overall curve.  It is about $5\%$ at 
$\sqrt{S}=630 \, {\rm GeV}$ and about $3 \%$ at $\sqrt{S}=1800\, {\rm GeV}$.
The order of magnitude of the result is reasonable, although 
higher than the roughly $1\%$ found at the  Tevatron
using track or tower multiplicities.   How much of this difference is due to
our new definition of the gap and how much to the lack of a nonperturbative
survival probability remains to be explored.
The sharp upturn that we observe below $1 \, {\rm{GeV}}$ is due to the 
divergence of the perturbative running coupling 
at $Q_c=\Lambda$; nonperturbative effects will attenuate this rise.     

Previous analysis of rapidity gaps in dijet events has tended to
emphasize either the short-distance \cite{Bjork,DDucaT} or long-distance
\cite{BuchHeb,eboli,zepp} aspects of the problem.
(The role of Sudakov logarithms in double-rapidity gap events
has been discussed in \cite{MRK}.)  Here, we have argued
that by factorizing short- and long-distance effects, we may treat both
dependences systematically.  In our formalism, the mixing of color
states begins at short distances precisely with two-gluon exchange \cite{Bjork,DDucaT},
summarized through the anomalous dimension $\Gamma_S$, while long-distance color
(``bleaching") effects \cite{BuchHeb,eboli,zepp} follow the evolution of the different color components
between the short-distance scale $\sqrt{-\hat{t}}$ and the long-distance
scale $Q_c$.  

As we have observed above, our formalism
does not include a nonperturbative survival probability \cite{Bjork,GLM} associated
with the interaction of spectator partons.   Clearly, a full phenomenological
analysis will also require the inclusion of processes involving
gluons (including $q\bar q\rightarrow gg$) and sea quarks.  The treatment
of gluon-gluon scattering \cite{eboli,zepp} should be particularly interesting \cite{KOS2}.
Nevertheless, we believe that the basic features shown in the 
valence-quark analysis outlined above will appear as well in
a more complete discussion.  A calorimetric
analysis of dijet rapidity gap events, if possible experimentally, could shed valuable light
on the dynamics of QCD.

\subsection*{Acknowledgments}
We are indebted to Jack Smith for his valuable advice 
in the implementation of the numerical simulation.  This work was
supported in part by the National Science Foundation, grant PHY9722101.

\end{document}